\documentclass[doublecol]{epl2} 
\usepackage{newtxtext}
\usepackage{graphicx} 
\usepackage{amsmath} 
\usepackage{amsfonts}
\usepackage{mathtools}
\usepackage{placeins}
\usepackage{verbatim}
\usepackage[dvipsnames]{xcolor}
\usepackage{pdfpages}
\newcommand{\includecomment}[1]{}
\newcommand{\Cl}{\mathrm{Cl}}
\newcommand{\Clp}{\mathrm{Cl}^{+}}
\newcommand{\SU}{\mathrm{SU}}

\title{Geometry and relaxation dynamics of nematic loops}
\shorttitle{Geometry and relaxation dynamics of nematics loops}

\author{F. Aprile\inst{1} \and  A. J. H. Houston\inst{2}  \and G. Gonnella\inst{1} \and D. Marenduzzo\inst{3} \and T.~N. Shendruk\inst{3} \and G. Negro\inst{3}}
\shortauthor{F. Aprile \etal}

\institute{                    
  \inst{1} Dipartimento Interateneo di Fisica, Università degli Studi di Bari and INFN, Sezione di Bari,\\ via Amendola 173, Bari, I-70126, Italy\\
  \inst{2} School of Mathematics and Statistics, University Place, Glasgow, G12 8QQ, UK\\
  \inst{3} SUPA, School of Physics and Astronomy, University of Edinburgh, Edinburgh EH9 3FD, UK
}
 
\abstract{Disclination lines in three-dimensional nematic liquid crystals generically form closed loops whose topology is classified by homotopy theory. While this classification successfully captures global topological features
it does not encode the geometry of the defect profile along the loop, which can strongly influence defect dynamics. Here, we propose a geometric description of nematic disclination loops using the Clifford algebra $\mathrm{Cl}(3,0)$. This approach naturally captures the geometry of the local defect profile, as well as changes along the loop, which is mathematically a SU(2) holonomy. Simulations of the dynamics of defect loops with specified geometries embedded in nematic liquid crystals demonstrate that loops nucleate the growth of ``topological blobs'' of defects, which later dissipate leaving uniform nematic textures. Self-twist of the defect profile leads to nucleation of additional linking disclination lines, with a simple arithmetic relation between total self-twist and linking number. In contrast, loops with an even number of discrete profile transitions generate patterns with threading between loops, but no linking. 
These results establish a direct connection between the geometric holonomy of a disclination loop and its subsequent evolution, and may be extendable to more complex order parameter manifolds, such as cholesterics or smectics.
}
\begin{document}

\maketitle

\section{Introduction}

Liquid crystals are paradigmatic systems with long-range orientational order and a rich defect phenomenology~\cite{degennes1993}. Beyond their technological relevance in display devices, they provide an ideal system to study topological defects in ordered media~\cite{mermin1979}. In nematics, the ordered field is a director $\mathbf{n}$, a unit vector with head--tail symmetry with ground-state manifold $S^2/\mathbb{Z}_2$. As a result, the fundamental group 
$\pi_1(S^2/\mathbb{Z}_2)=\mathbb{Z}_2$ classifies straight line defects (disclinations) into two classes: topologically trivial and non-trivial~\cite{mermin1979}. 

While in 2D defects are pointlike, disclinations generically form closed loops in 3D ~\cite{Mkaddem2000,Kos2020},  characterized by a  richer topology than that of an isolated straight line~\cite{selinger2024introduction,alexander2012}. In particular, loop configurations are classified by $\mathbb{Z}_4$~\cite{janich1987,vcopar2014}, distinguishing two classes of unpierced loops from another two topological classes describing loops threaded by other disclinations. While this topological classification constrains global transformations and reconnection events, it provides only a coarse-grained description of loop structure. 

The work of J\"{a}nich considers the homotopy classes of the nematic on a toroidal neighbourhood of a defect loop \cite{janich1987}. Still richer topological information has been found by considering the entire nematic texture, with the topologically distinct director fields corresponding to elements of a twisted cohomology group of the complement of the defect link \cite{machon2014knotted,machon2016global,machon2019topology}, with at least a partial explanation given in terms of umbilic lines \cite{machon2016umbilic}.

Recent work has emphasised that the geometry of disclination loops --- namely, the detailed structure of the defect profile and its transport along the loop --- contains physically relevant information beyond topology~\cite{long2021,head2024b}. Locally, the defect profile in a plane perpendicular to the line can be represented by a triad, or ``flagpole'', description~\cite{vcopar2013,vcopar2014,kos2022}. The pole encodes the axis about which the director rotates by $\pi$, while an in-plane phase specifies the orientation of the flag. Together, these degrees of freedom define an element of $S^3 \simeq \mathrm{SU}(2)$, or equivalently a unit quaternion. As this triad is transported along the loop, it also accumulates non-trivial SU(2) holonomy~\cite{vcopar2014}, which refines the $\mathbb{Z}_4$ classification, capturing geometric information, such as self-twist and profile transitions from comet to trefoil that have topological charge $+1/2$ and $-1/2$ on the plane~\cite{degennes1993}. Geometrical descriptions such as these are useful to understand the physics of disclinations in greater depth than via a topological classification alone, as seen for instance in cholesterics, where topological rules alone cannot explain physically observed defect recombinations~\cite{beller2014,johnson2025}.

This work proposes to describe this geometric structure using the three-dimensional Clifford algebra $\mathrm{Cl}(3,0)$~\cite{lounesto2001,hestenes2003,head2024b,johnson2025}. This framework naturally unifies a local description of the defect profile with its global transport properties along the loop. It also allows the loop geometry to be decomposed into algebraically transparent components. In particular, self-twist (torsional transport of the profile) and discrete profile changes appear as distinct contributions to the SU(2) holonomy that describes the intrinsic variation of the triad along the disclination loop. This representation may be useful for others who wish to develop geometric representations of 3D disclination lines in liquid crystals.

We further investigate how these geometric features influence defect dynamics using continuum simulations of a nematic liquid crystal. Studying the dynamics of topological defects following the insertion of a circular disclination loop with different geometries into an otherwise uniform nematic liquid crystal reveals that different geometries lead to qualitatively distinct transient behaviours. A globally self-twisted loop nucleates additional linking disclination lines, with the number of threads set by a simple arithmetic relation with the total self-twist. By contrast, loops containing an even number of profile transitions create a localised pattern of defect loops which are unlinked but thread each other, resembling threading polymer rings~\cite{michieletto2014,smerk2020}. In all cases, the inserted loop nucleates a ``topological blob'' of defects, which is reminiscent of the turbulent blob accompanying the insertion of vortex rings in fluids~\cite{matsuzawa2023}.

These results demonstrate that the geometry of a disclination loop is not merely a mathematical refinement of topology, but a physically predictive quantity with detectable signatures, for instance in the dynamics observed in the medium. Indeed, to our knowledge, a direct link between SU(2) holonomy and subsequent defect nucleation dynamics has not previously been demonstrated. The Clifford algebra formulation provides a systematic way to encode this information, and may prove useful more broadly to study other defects in ordered~\cite{henrich2010,henrich2011} and active media~\cite{binysh2020,houston2022,Duclos2020,digregorio2024}.

\section{Clifford algebra description of constant wedge disclination loops}

We start by considering a defect loop, namely an unknotted circular disclination line. A disclination can be characterized by its local profile, namely the configuration of the nematic field near the defect core on a plane perpendicular to the tangent vector at a given point along the loop.  

\begin{figure}[ht]
    \centering
    \includegraphics[width=\columnwidth]{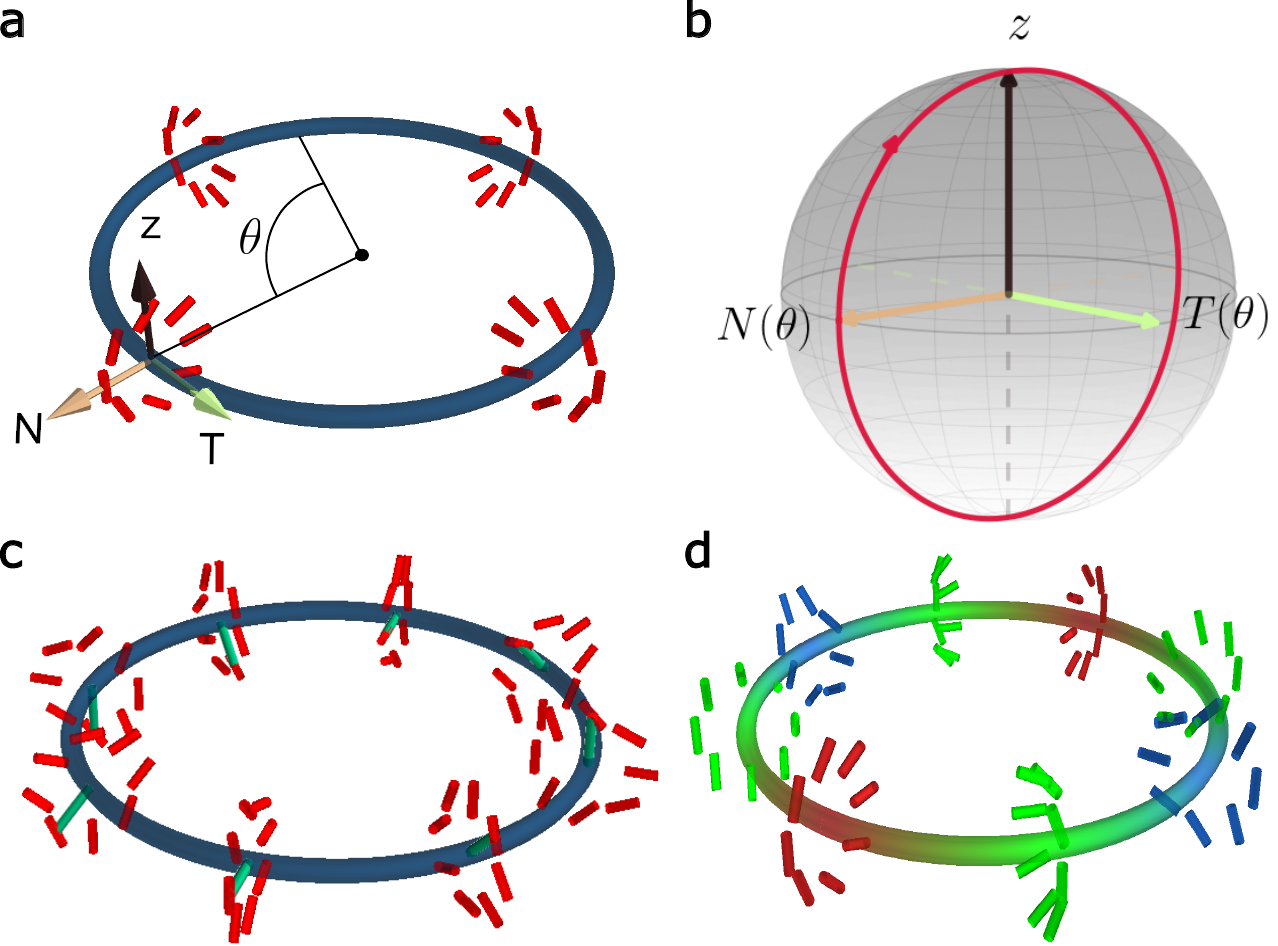}
    \caption{{\textbf{Geometry of disclination loops.}} (a) Defect loop parameterised by the angle \(\theta\) with a constant \(+1/2\) local wedge profile and the associated local basis formed by \(\mathbf{T}(\theta)\), tangent to the disclination, \(\mathbf{N}(\theta)\), radially outward, and  \(\mathbf{z}\). 
    (b) Corresponding circle in parameter space at fixed \(\theta\) going through the antipodal points \(z=\pm1\).
    (c) Defect loop with a constant \(-1/2\) local wedge profile and a self-twist \(\tau(\theta)=\theta\) around the tangent axis \(\mathbf{T}(\theta)\) (green rods).
    (d) Plus-minus-plus-minus (PMPM) loop, characterized by four changes of the local profile from \(+1/2\) local wedge to \(-1/2\) local wedge. Equivalently, this is described as a rotation of the profile around the \(\mathbf{z}\) axis with \(\beta(\theta)=-2\theta\).
    Trefoil (\(-1/2\)) profiles are shown in blue, comet (\(+1/2\)) profiles in red, and intermediate twist profiles in green.
    }
    \label{fig1}
\end{figure}

To describe this geometry, consider a torus surrounding the defect loop, parametrised by two angles \(\theta\) and \(\phi\), describing respectively the longitude and meridian of the torus. We introduce a local Frenet–Serret frame with constant binormal $\mathbf{z}=(0,0,1)$, tangent $\mathbf{T}(\theta)=(-\sin\theta,\cos\theta,0)$ and normal $\mathbf{N}(\theta)=(\cos\theta,\sin\theta,0)$, as befits a planar curve (Fig.~\ref{fig1}.a): the loop lies in the plane $z=0$, $\mathbf{T}(\theta)$ is the local tangent to the disclination line, and $\mathbf{N}(\theta)$ is orthogonal to both, pointing radially outward for a perfectly circular loop.

We now fix \(\theta\) and examine the local configuration of the nematic profile around the selected small circle. This local configuration can be naturally described by a circle in parameter space- namely the 2-sphere with antipodal points identified. (Fig.~\ref{fig1}b). The circle  encodes the directions swept out by the nematic director along the selected small circle.
{For example, one can consider a constant \(+1/2\) local wedge profile, such as the one shown in Fig.~\ref{fig1}a. In this case, for each value of \(\theta\), namely for each small loop on the torus, the local profile is a \(+1/2\) wedge profile. In parameter space, this profile is described by a circle, shown in red in Fig.~\ref{fig1}b, lying in the plane spanned by \(\mathbf{z}\) and \(\mathbf{N}(\theta)\).}
This circle can be parametrized by the vector
\begin{equation}
\begin{split}
    \mathbf{r}(\theta,\phi)
    &= \mathbf{z}\cos\phi + \mathbf{N}(\theta)\sin\phi \\
    &= \mathbf{z}\cos\phi + (\mathbf{x}\cos\theta + \mathbf{y}\sin\theta)\sin\phi .
\end{split}
\end{equation}

{To describe the the loop in terms of Cliffor algebra, we consider the three-dimensional  Clifford algebra \(\mathrm{Cl}(3,0)\)~\cite{lounesto2001,johnson2025,head2024b}.
The algebra is generated by three elements \(e_1, e_2, e_3\), satisfying
\(e_i^2 = 1\) for \(i=1,2,3\), and \(e_i e_j = -e_j e_i\) for \(i \neq j\).
Spinors, in the Clifford representation, are elements of the even subalgebra \(\mathrm{Cl}^+(3,0)\), generated by scalars and bivectors \(e_i e_j\). We denote the bivector \(e_{ij} \equiv e_i e_j\) and pseudoscalar \(e_{ijk}\equiv e_i e_j e_k\), and it is useful to introduce an explicit matrix representation. A convenient choice is given by the identity matrix and Pauli matrices\cite{SI},

which satisfies the algebraic conditions for the generators after identifying \(e_i \leftrightarrow \sigma_i\) and \(1 \leftrightarrow \mathbf{1}\).}

The defect spinor is constructed by considering the eigenvectors of
\begin{equation}
    \mathbf{r}(\theta,\phi)\cdot \boldsymbol{\sigma}
    =
    \begin{pmatrix}
    \cos\phi & e^{-i\theta}\sin\phi \\
    e^{i\theta}\sin\phi & -\cos\phi
    \end{pmatrix},
\end{equation}
which correspond to the eigenvalues \(\lambda = \pm 1\). 
Juxtaposing the two eigenvectors produces the Clifford element 
\begin{equation}\label{wedgePlus}
\begin{aligned}
    \Psi(\theta,\phi) &= 
    \begin{pmatrix}
    \cos(\phi/2) & -e^{-i\theta}\sin(\phi/2) \\
    e^{i\theta}\sin(\phi/2) & \cos(\phi/2)
    \end{pmatrix} 
    \\ 
    &= \mathbf{1}\cos(\phi/2)
    + e_{23}\sin\theta \sin(\phi/2) \\
    &\qquad - e_{31}\cos\theta \sin(\phi/2) ,
\end{aligned}
\end{equation}
which is an element of the even subalgebra \(\mathrm{Cl}^+(3,0)\). The element \(\Psi\) describes a defect loop with constant \(+1/2\) local wedge profile. 
A constant \(-1/2\) wedge profile is constructed by applying the transformation \(\phi \to -\phi\), which gives 
\begin{equation}\label{WedgeMinus}
\begin{split}
    \Xi(\theta,\phi) &=
    \mathbf{1}\cos(\phi/2)
    - e_{23}\sin\theta \sin(\phi/2) 
    \\
    &\qquad + e_{31}\cos\theta \sin(\phi/2).
\end{split}
\end{equation}

{The Clifford algebra $\Cl(3,0)$ provides a unified language in which the local profile is encoded as a spinor in the even subalgebra $\Clp(3,0) \simeq \SU(2)$. In addition, as we will further see in the following, rotations of the profile act multiplicatively as rotors, and
 the $\SU(2)$ holonomy of the profile along the loop is exposed as an algebraically transparent product of rotor factors. Geometric quantities such as self-twist and discrete profile transitions thereby appear as distinct,
composable contributions to a single object.}

\section{Generalisation of Clifford algebra description to varying defect profiles}

In the previous section, we obtained an explicit description of constant wedge \(\pm 1/2\) defect loops. We now generalize this construction to describe a generic circular defect loop as an element of a Clifford algebra. In general, the local profile can undergo a smooth rotation along the loop around a given axis. Three independent rotation axes can be naturally identified, for instance those defined by the local basis introduced previously. A rotation of the profile around \(\mathbf{T}(\theta)\) introduces a self-twist of the profile; a rotation around \(\mathbf z\) induces a transition between local \(+1/2\) and \(-1/2\) wedge profiles; while a rotation around \(\mathbf N(\theta)\) induces a transition between local wedge and local twist profiles. In this section, we implement these rotations separately on the constant wedge profiles defined in Eqs.~\eqref{wedgePlus} and \eqref{WedgeMinus}. To this end, we exploit the properties of rotations in Clifford algebras, which allow us to perform the transformation through multiplication by a rotor.  In Clifford algebra, an anticlockwise rotation of angle \(\alpha\) around a unit vector \(\mathbf{n}\) is implemented by
\begin{equation}\label{GeneralRotation}
    R_{\mathbf{n}}(\alpha)
    =
    \mathbf{1}\cos(\alpha/2)
    - e_{123}\mathbf{n}\sin(\alpha/2).
\end{equation}
We apply rotations (Eq.~\eqref{GeneralRotation}) about the three local bases to describe loops with varying profiles. 

Firstly, \includecomment{torsion} self-twist around the tangent vector \(\mathbf{T}(\theta)\) by an angle $\tau(\theta)$ is given by the rotation
\begin{equation}\label{twistRotation}
\begin{split}
    R_{\mathbf{T}}(\tau(\theta))
    &=
    \mathbf{1}\cos\!\left(\frac{\tau(\theta)}{2}\right)
    - e_{123}\mathbf{T}(\theta)\sin\!\left(\frac{\tau(\theta)}{2}\right) \\
    &=
    \mathbf{1}\cos\!\left(\frac{\tau(\theta)}{2}\right)
    + e_{23}\sin\theta \sin\!\left(\frac{\tau(\theta)}{2}\right) \\
    &\qquad - e_{31}\cos\theta \sin\!\left(\frac{\tau(\theta)}{2}\right).
\end{split}
\end{equation}
Applying the operator \(R_{\mathbf{T}}\) (Eq.~\eqref{twistRotation}) to the \(-1/2\) disclination profile (Eq.~\eqref{WedgeMinus}) gives the spinor associated with a self-twisted loop defect with a constant \(-1/2\) local wedge profile (Fig.~\ref{fig1}) to be
\begin{equation}\label{twistApplied}
\begin{split}
    R_{\mathbf{T}}(\tau(\theta)) \ \Xi(\theta, \phi)
    &=
    \mathbf{1}\cos\!\left(\frac{\phi-\tau(\theta)}{2}\right) \\
    &\qquad - e_{23}\sin\theta \sin\!\left(\frac{\phi-\tau(\theta)}{2}\right) \\
    &\qquad\quad + e_{31}\cos\theta \sin\!\left(\frac{\phi-\tau(\theta)}{2}\right),
\end{split}
\end{equation}
This shows that \includecomment{torsion}self-twist introduces a running phase offset. 
Owing to the intrinsic trefoil symmetry of the loop, linear self-twist \(\tau(\theta) = m\theta\) is restricted to twist rates \(m=n/3\) for \(n\in\mathbb{Z}\), i.e. only integer multiples of \(1/3\) are admissible. We refer to the integer \(n\) as the ``twist rate index''. 
Loops for which the defect profile twists may be referred to as having  torsional holonomy. 

Next let us consider rotations about the \(\mathbf{z}\) axis. 
Rotation by an angle \(\beta(\theta)\) around the \(\mathbf{z}\) axis is given by
\begin{equation} \label{TorsionSpinor}
    R_{\mathbf{z}}(\beta(\theta))
    =
    \mathbf{1}\cos(\beta(\theta)/2)
    - e_{12}\sin(\beta(\theta)/2). 
\end{equation}
Applying the rotation around \(\mathbf{z}\) (Eq.~\eqref{TorsionSpinor}) to the \(-1/2\) constant wedge profile (Eq.~\eqref{WedgeMinus}) produces the profile 
\begin{equation} \label{rotationZ}
\begin{split}
    R_{\mathbf{z}}(\beta(\theta)) \ \Xi(\theta,\phi) &= \mathbf{1}\cos{(\phi/2)}\cos{(\beta(\theta)/2)}\\
    &\quad +e_{23}\sin{(\phi/2)}\sin{(\theta + \beta(\theta)/2)} \\ 
    &\quad +e_{31}\sin{(\phi/2)}\cos{(\theta+\beta(\theta)/2)} \\
    &\quad -e_{12}\cos{(\phi/2)}\sin{(\beta(\theta)/2)} . 
\end{split}
\end{equation}
By selecting different functions for \(\beta(\theta)\), \(R_{\mathbf{z}}\Xi\) (Eq.~\eqref{rotationZ}) allows us to describe disclinations for which the local wedge profile changes from \(+1/2\) wedge to \(-1/2\) wedge. 
For example, choosing \(\beta(\theta) =- 2\theta\) gives the plus-minus-plus-minus (PMPM) profile shown in Fig.~\ref{fig1}.d, with the associated spinor
\begin{equation}
\begin{split}
    \Psi_\mathrm{PMPM}(\theta,\phi)
    &=\mathbf{1}\cos{(\phi/2)}\cos{(\theta)} + e_{31}\sin{(\phi/2)}\\ 
    &\qquad +e_{12}\cos{(\phi/2)}\sin{(\theta)} .
\end{split}
\end{equation}

Having considered the first two axes in the local basis, we finally turn our attention to rotations about \(\mathbf{N}(\theta)\). 
A rotation around \(\mathbf{N}(\theta)\) by an angle $\gamma(\theta)$ is implemented by \cite{SI}
\begin{equation}
\begin{split}
    R_{\mathbf{N}}(\gamma(\theta))
    &=
    \mathbf{1}\cos(\gamma(\theta)/2)
    - e_{23}\cos\theta \sin(\gamma(\theta)/2) \\
    &\quad - e_{31}\sin\theta \sin(\gamma(\theta)/2).
\end{split}
\end{equation}

Applying \(R_{\mathbf{N}}\) to Eq.~\eqref{wedgePlus} gives
\begin{equation}
\begin{split}
    R_{\mathbf{N}}(\gamma) \ \Psi(\theta,\phi)
    &=
    \mathbf{1}\cos(\gamma/2)\cos{(\phi/2)} \\
    &\quad +e_{23}[\sin\theta \cos(\gamma/2)\sin{(\phi/2)} \\ 
    &\qquad -\cos\theta \sin(\gamma/2)\cos{(\phi/2)}] \\
    &\quad - e_{31}[\sin\theta \sin(\gamma/2)\cos{(\phi/2)} \\
    &\qquad +\cos\theta \cos(\gamma/2)\sin{(\phi/2)}] \\
    &\quad -e_{12}\sin{(\gamma/2)}\sin{(\phi/2)}. 
\end{split}
\end{equation}

{The element $\Psi \in \mathrm{Cl}^{+}(3,0)$ compactly encodes the local profile and its transport.  To go back to a frame-like geometric description,  we can project $\Psi$ back to a vector via the Hopf-like  map ~\cite{lounesto2001} 
\begin{equation}
  \Psi \;\longmapsto\; \Psi\, e_{3}\, \widetilde{\Psi},\label{hopf}
\end{equation}
which inverts the spinorial construction of Eq.~(6). Here \( \ \widetilde{\cdot} \ \) represents the operation of changing the sign of the bivector part of the element.}

As an example, applying the mapping in Eq.~\eqref{hopf} to the element from Eq.~\eqref{twistApplied} results in the vector
\begin{equation}
\begin{split}
    \mathbf{r}_{\tau}(\theta,\phi) &= R_\mathbf{T}\Xi e_3 \widetilde{\Xi}\widetilde{R}_\mathbf{T} \\
    &=-\sin{(\phi-\tau(\theta))}(e_1 \cos{\theta+e_2\sin{\theta}}) \\
    &\qquad +e_3\cos{(\phi-\tau(\theta))} . 
\end{split}
\end{equation}
{In particular, the \includecomment{torsion} self-twist does not change the plane in which the vector lies. The vectorial framework is particularly well suited for extracting geometrical quantities, such as the winding vector \(\boldsymbol{\Omega}(\theta)\) \cite{johnson2025,schimming2022,schimming2023}, defined as the vector perpendicular to \(\mathbf{r}(\theta,\phi)\) for all values of \(\phi\). Therefore,
\begin{equation}
    \langle \mathbf{r}_\tau,\boldsymbol{\Omega} \rangle=0
    \quad \Longrightarrow \quad
    \boldsymbol{\Omega}(\theta)
    = -e_1\sin{\theta}+e_2\cos{\theta}.
\end{equation}}

Another example is to apply the mapping of Eq.~\eqref{hopf} to the PMPM loop. The transformed profile is then
\begin{equation}
\begin{split}
    \mathbf{r}_\mathrm{PMPM}(\theta,\phi) &= \Psi_{PMPM}e_3\widetilde{\Psi}_{PMPM} \\
    &= -\sin{\phi}(e_1\cos{\theta}+e_2\sin{\theta}) +e_3\cos{\phi}
\end{split}
\end{equation}
and the \(\mathbf{\Omega}\) vector is given by
\begin{equation}
    \mathbf{\Omega}=e_1\sin\theta + e_2 \cos{\theta} . 
\end{equation}
{Equations (10), (12) and (15), demonstrate that the Clifford framework provides a unified algebraic representation of circular disclination loops with arbitrary geometrical features. The framework, though,  says nothing about how that geometry evolves in the bulk, where elastic response of the liquid crystal  can nucleate additional defect structures. In the next sections we will  probe how the imposed $\SU(2)$ holonomy translates into observable dynamics though numerical simulations. }

\section{Relaxation dynamics}
We now ask how the geometric features discussed so far influence defect dynamics. To this aim we perform continuum simulations of a nematic liquid crystal, characterizing the dynamics of topological defects following the insertion of a circular disclination loop with different geometries into an otherwise uniform nematic liquid crystal.

As in previous studies~\cite{carenza2019,SI},
we model a three-dimensional nematic liquid crystal in terms of the order parameter tensor $\mathbf{Q}(\mathbf{r},t)$ and the  velocity field $\mathbf{v}(\mathbf{r},t)$,  
with the thermodynamic properties described by a Landau–de Gennes free energy~\cite{degennes1993},  under a one-constant approximation~\cite{carenza2019}. The dynamics is governed by the Beris–Edwards equations~\cite{beris} for  $\mathbf{Q}$, and the incompressible Navier–Stokes equations for the fluid velocity $\mathbf{v}$ (see~\cite{SI} for more details). 
The equations are solved using a three-dimensional hybrid lattice Boltzmann method~\cite{carenza2019,SI}. 

Simulations are in 3D on a cubic lattice with periodic boundary conditions, for which topological charge conservation requires the total topological charge of defects in the system be equal to \(0\) modulo \(2\)~\cite{mermin1979}. The configurations investigated are initialized from a defect torus with large and small radii \(R_1\) and \(R_2\), respectively, centered in the middle of the simulation domain. 

In all cases, outside the toroidal region the system is initialized in the isotropic phase, corresponding to \(\mathbf{Q} = \mathbf{0}\), with a small random noise added. More details about numerical simualtions are given in~\cite{SI}.

\begin{figure}[t!]
    \centering
    \includegraphics[width=1.0\columnwidth]{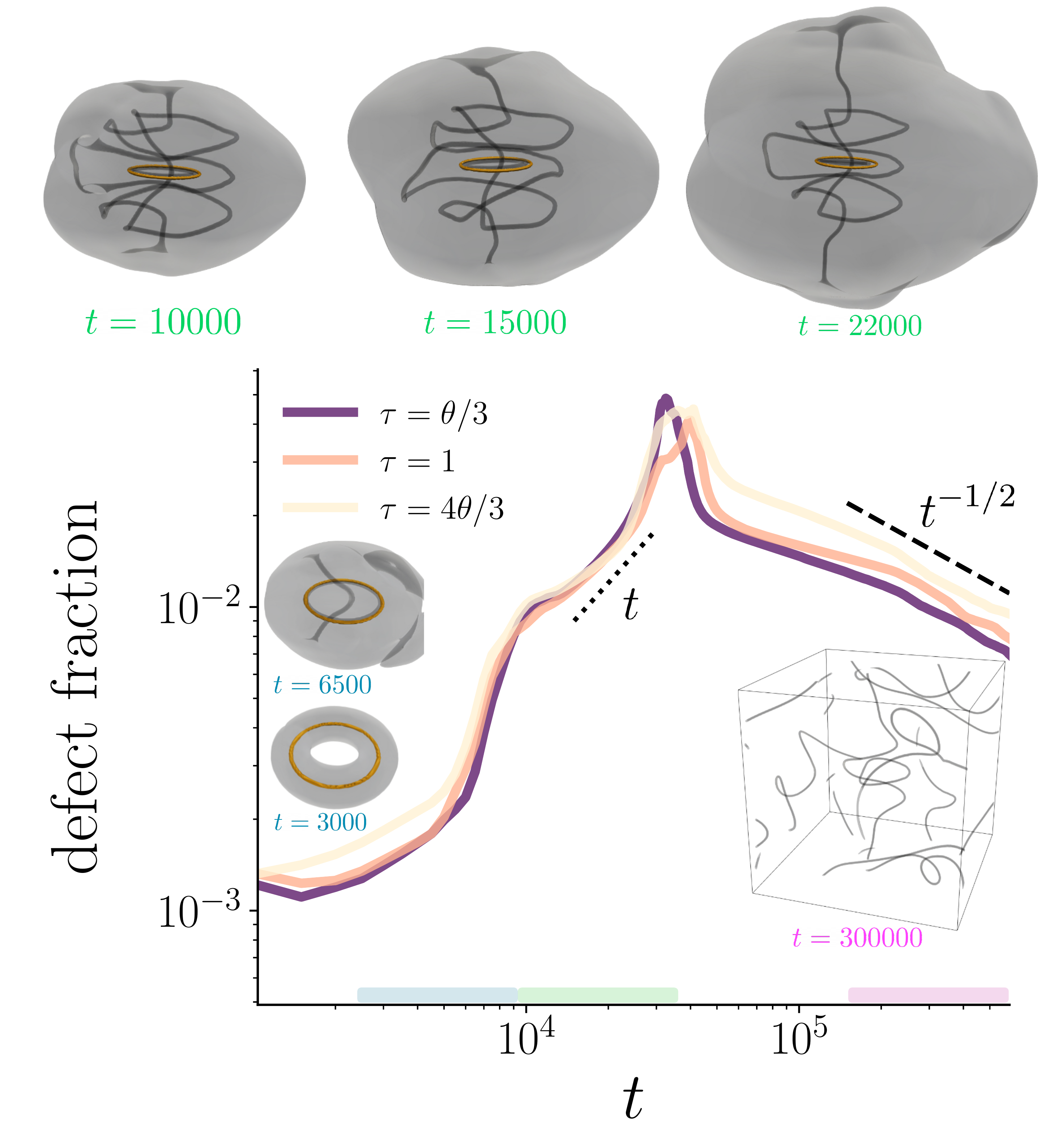}
    \caption{{\textbf{Relaxation of self-twisted wedge loops}.} Defect fraction over time for different \includecomment{torsion} self-twist $\tau=n\theta/3$. Three twist rates corresponding to $n=1, 3$ and $4$ are shown.  
    Three time intervals of early (blue), intermediate (green) and late (pink) are identified. 
    Insets show snapshots of $q$ isosurfaces for the case \(\tau=\theta/3\) at early times ($t=3000$ and $6500$; left), intermediate times ($t=10000, \ 15000, \ 22000$; top) and late times ($t=300000$; right). Initial loop is represented in orange.}
    \label{fig2}
\end{figure}

\begin{figure*}[ht]
    \centering
    \includegraphics[width=2.0\columnwidth]{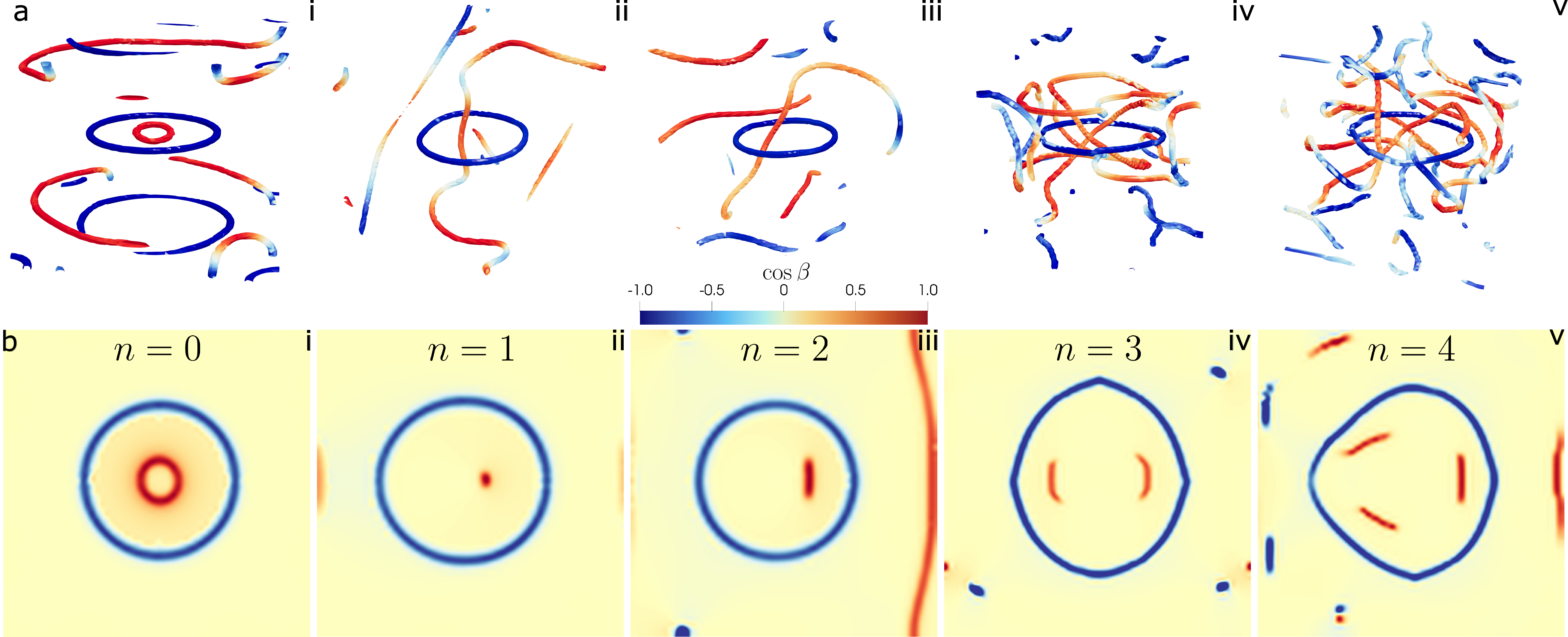}
    \caption{\textbf{Linked configurations for self-twisted wedge loops.}
    (a) Defect configurations in the intermediate time interval for increasing \includecomment{torsion} self-twist  $\tau = n\theta/3$. Disclinations identified as isosurfaces of $s$ the modulus of the disclination tensor and coloured by $\cos\beta$.
    (b) Cross-sectional maps of $s\cos\beta$ in the plane defined by $\mathbf{z}$ for the same configurations. 
    (panel i) $n=0$, (ii) $n = 1$, (iii) $n = 2$, (iv) $n=3$, and (v) $n=4$. 
    The configurations shown in panels b.i and b.ii for $t=10000$; while panels b.iii, b.iv and b.v for $t=20000$.}
\label{fig3}
\end{figure*}

\subsection{Relaxation of \(-1/2\) wedge profiles}

We start by considering initial conditions consisting of a charged defect loop with a constant \(-1/2\) wedge profile and a \includecomment{torsion} self-twist \(\tau(\theta)=n\theta/3\) with \(n \in \mathbb{Z}\). In Clifford algebra, this corresponds to the spinor given by Eq.~\eqref{twistApplied} in which the self-twist creates a running phase proportional to the twist rate index \(n\). {The insets of Fig.~\ref{fig2} show the time evolution of the isosurface $q=0.3$,  starting from  an initial condition consisting of a defect loop with a constant \(-1/2\) wedge profile and \includecomment{torsion} \(\tau(\theta)=\theta/3\), i.e. \(n=1\).}

At the earliest times (Fig.~\ref{fig2}; \(t=3000\) inset), the disclination retains the initial loop structure. However, soon after additional defect structures form due to the interplay between topological constraints and energy minimisation, inducing the formation of a straight disclination  line  that goes through the centre of the loop (Fig.~\ref{fig2}; \(t=6500\) inset), and will eventually (see later) span the periodic boundaries.
Because the straight disclination spans the periodic boundaries, it constitutes a loop, and so it and the initial loop together form a ``linked loop'' pair. 

In addition to the linked loop, a ``topological blob'', with a soliton-like nature is observed to arise at early times (region enclosed by the gray surface in Fig.~\ref{fig2}; \(t=6500\) inset). 
We track the evolution of the size of this topological blob by measuring the defect fraction as a function of time $t$. The defect fraction is computed as the fraction of grid points where the maximum eigenvalue $q$ is within the range $0.3 \leq q \leq 0.33$. The defect fraction grows superlinearly at early times (time interval highlighted in blue) and then approximately linearly at intermediate times (green time interval). The defect fraction in these early and intermediate time intervals exhibits little apparent dependence on the \includecomment{torsion} self-twist of the defect profile. The initial fast growth of the topological blob resembles the spreading dynamics of a turbulent blob, created by the collision of vortex rings, into quiescent flow observed in~\cite{matsuzawa2023,matsuzawa2026}. The nematic loop disclination system can be viewed as the elastic analogue of this, as the dynamics is governed only by thermodynamic torques rather than hydrodynamic flow as in~\cite{matsuzawa2023,matsuzawa2026}.

In the intermediate time interval highlighted in green, the topological blob expands while the connecting disclination lines attached to its surface elongate (Fig.~\ref{fig2}; $t=10000-20000$ insets). During this process, the internal topological configuration remains preserved, as shown in the top insets.

The intermediate time interval ends with a peak in the defect fraction (Fig.~\ref{fig2}). 
This peak corresponds to the blob reaching the domain boundaries (not shown). From this point onward, the system relaxes toward a defect-free nematic equilibrium state, compatible with the topological constraints and minimizing the free energy.
Interestingly, at sufficiently late times (time interval highlighted in pink), the defect fraction decays with time as $t^{-1/2}$ until defect-free nematic equilibrium is achieved.
To characterize how the imposed geometric holonomy controls the transient defect dynamics, we focus on the intermediate time window ($t=10000$ to $20000$ highlighted in green in Fig.~\ref{fig2}), in which the topological blob expands while the number and character of disclination lines remain unchanged.
In  this regime, varying the \includecomment{torsion} twist rate index $n$ leads to the formation of linked structures originating from the initial loop. 
Representative configurations with disclination lines are shown in Fig.~\ref{fig3}.a.i--v for
twist rates \includecomment{\(\nu=0,1/3,2/3,1\), and \(4/3\)}\(n=0,1,2,3\), and \(4\), respectively. 
The disclination lines are coloured by $\cos\beta$, with trefoil (\(-1/2\)) profiles corresponding to $\cos\beta\approx-1$ in blue, comet (\(+1/2\)) profiles to $\cos\beta\approx+1$ in red, and twist profiles to $\cos\beta\approx0$ in yellow.
When the twist rate is equal to \(n=0\) \includecomment{zero torsion}, a small loop forms close to, and approximately concentric with, the original loop, but no linked loop is generated. 
This can be seen clearly in both 3D (Fig.~\ref{fig3}.a.i) and the cross section (Fig.~\ref{fig3}.b.i). 

The cross-sectional maps (Fig.~\ref{fig3}.b) show the signed defect-density field \(\mathrm{Tr}\,\mathbf{D} = s\cos\beta\)\cite{SI}.  Since \(s\) is peaked at the defect cores, these maps provide a two-dimensional projection of both the defect position and its local profile. The bright blue outer circular features in Fig.~\ref{fig3}.b correspond to the original loops, while the additional localized regions inside the loop correspond to the nucleated linked-loop segments. 
In Fig.~\ref{fig3}.b.i, the red ring corresponds to the unlinked concentric $+1/2$ wedge-type ring in plane with the original loop $-1/2$ loop. 

As the \includecomment{torsion} twist rate index is increased from $n=0$ to $n=1$, linked loops begin to appear, in which a relatively straight disclination line (that loops over the periodic boundaries) passes through the initial loop forming a linked loop (Fig.~\ref{fig3}.a.ii and \ref{fig3}.b.ii). 
As the twist rate is increased to $n=2$, 
the topology of the configuration is still a single linked loop pair (Fig.~\ref{fig3}.b.iii). 
However, for \includecomment{\(\nu=1\)}\(n=3\), two linked loops are present with two nucleated loops linked through the initial loop (Fig.~\ref{fig3}.a.iv), which can be seen most clearly in the cross section (Fig.~\ref{fig3}.b.iv). 
Likewise, for \includecomment{(\(\nu=4/3\)}\(n=4\), three linked loops are produced (Fig.~\ref{fig3}.a.v-\ref{fig3}.b.v). The same counting persists over the full set of simulated twist rates (see \cite{SI} for additional cases) suggesting a relation between number of links and the \includecomment{torsion}twist rate index of the form
\begin{equation}\label{NLinks}
    \ell = \left\lfloor \frac{2n+1}{3} \right\rfloor .
\end{equation}
Here, \(\ell\) denotes the number of links of the defect loop, \includecomment{\(n\)} \(n\) is the \includecomment{torsion}twist rate index for self-twist $\tau=n\theta/3$ and \(\lfloor \cdot \rfloor\) denotes the floor function. 
The simulations are found to agree well with Eq.~\eqref{NLinks} even for higher values of \(n\)\cite{SI}, 
{showing that the self-twist of the initial profile,  is converted, during relaxation, into integer-quantised linking with newly nucleated disclination lines, highlighting a directly observable dynamical signature of  $\SU(2)$ holonomy.}

We now turn to the local defect character as \includecomment{torque} the twist rate increases.
In Fig.~\ref{fig3}.a.i-v, disclinations lines are 
colored according to the value of \( \cos\beta=\mathbf{T}\cdot\mathbf{\Omega} \), which reveals the local nature of the disclinations. 
For all $n$, the $-1/2$ profile of the initial loop is preserved at intermediate times, since $\cos\beta\approx-1$ at all points along the loop in each case. 
Next we consider the character of the disclinations forming linked loops with the initial loop. 

The newly formed defect segments in the central region
of the torus show a robust comet-like character (Fig.~\ref{fig3}.a.ii-v). This indicates that an initial loop with a \(-1/2\) wedge profile preferentially nucleates \(+1/2\)-like segments in its interior.
This behaviour is even clearer looking at the cross-sectional maps in Fig.~\ref{fig3}.b.ii-v. 
 As \includecomment{\(\nu\)}\(n\) is
increased, the number of such internal peaks increases as prescribed by Eq.~\eqref{NLinks}. 
Finally,  comet and trefoil profiles are predominant in all the analysed configurations, while local twist profiles are less frequent~\cite{SI} which is different from what is observed in 3D extensile active disclination loops~\cite{binysh2020,houston2022}.

\subsection{Relaxation of plus-minus profiles}

\begin{figure}[t]
    \centering
    \includegraphics[width=\columnwidth]{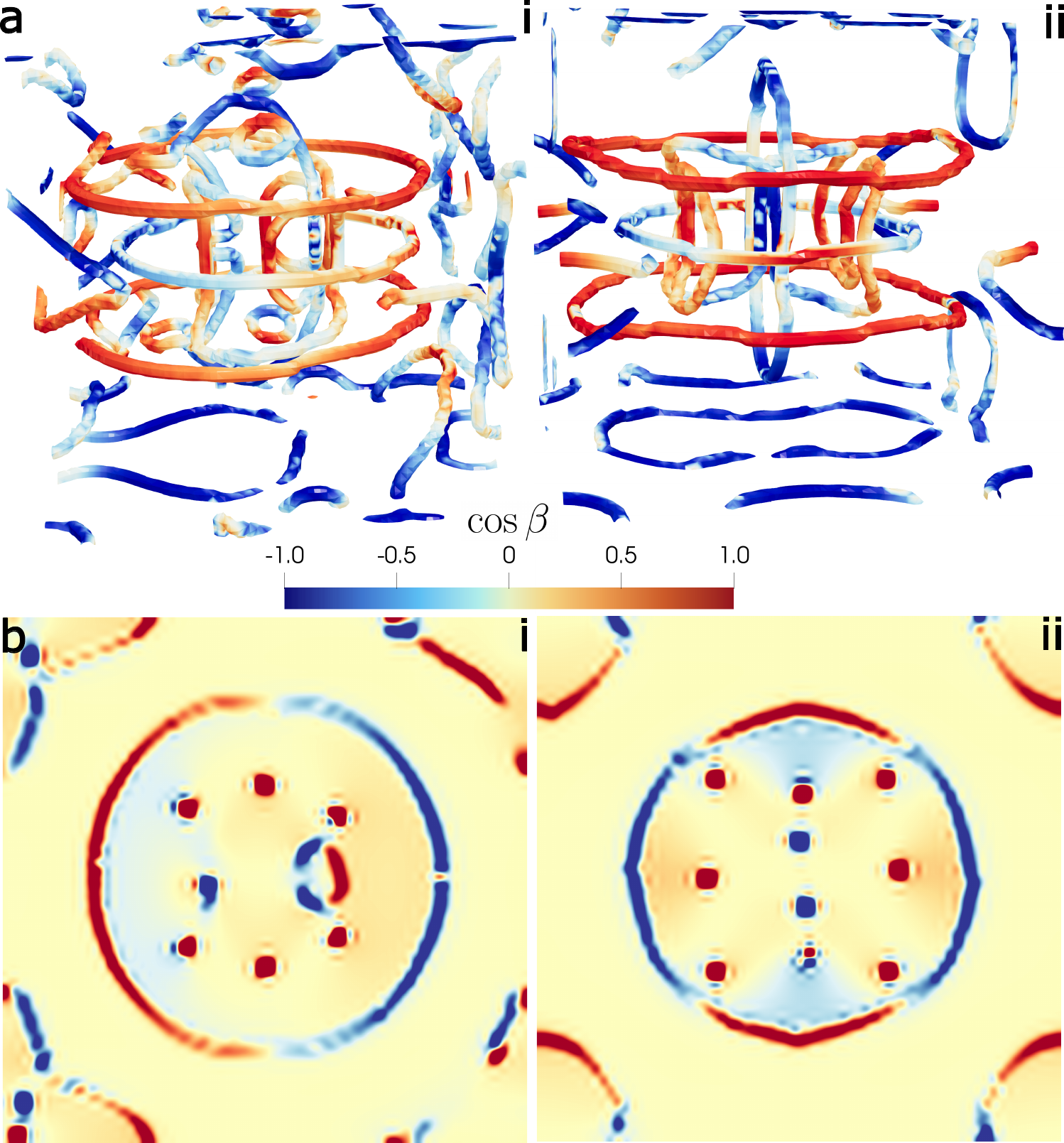} \caption{\textbf{Threaded  configurations for PM and PMPM loops.}
    (a) Defect configurations initialized with zero self-twist ($\tau_\mathrm{T}=0$). 
    The initial loop corresponds to a PM profile (panel i) and a PMPM profile (panel ii). 
    While no linked loops are formed since the imposed torsion vanishes, additional defect structures are nucleated and appear threaded (but not linked) with the initial loop. 
    (b) Cross-sectional mapping of the trace of the disclination tensor $s\cos\beta$. }
\label{fig5}
\end{figure}

We also considered plus-minus (PM) and PMPM initial profiles, shown in Fig.~\ref{fig5}a.i-ii, respectively. 
In both cases, the imposed 
self-twist is equal to $0$. This amounts to an effective total twist rate index of $n=0$. Given that the  twist rate is zero, substituting $n=0$ into Eq.~\eqref{NLinks} predicts that $\ell=0$, meaning no linked loops are expected to be generated. This is indeed what is observed in the simulations: unlike the \includecomment{torsion-dominated} self-twisted cases, the additional disclination structures that form for PM-type loops do not form loops linked with the original one. Instead, they appear as defect loops with the entire nucleated loop interpenetrating through the region enclosed by the initial loop (Fig.~\ref{fig5}.a.i and \ref{fig5}.a.ii). 
We refer to such interpenetrated but unlinked configurations as ``threaded'', in analogy to threadings associated with polymer melts~\cite{michieletto2014,smrek2016,smerk2020}. 

The difference between the PM and PMPM cases is nevertheless visible in the internal organisation of the nucleated defect loops, as is evident in the cross-sectional maps of $s\cos\beta=\mathrm{Tr}\,D$ (Fig.~\ref{fig5}.b). These maps provide a two-dimensional slice of both the position and local character of the nucleated threaded loops.
As in the \includecomment{torsion-dominated} self-twisted cases (Fig.~\ref{fig3}.b), comet-like segments are localised in the vicinity of the initial loop (Fig.~\ref{fig5}.b). 
{However, in contrast to the previous self-twisted cases, the region enclosed by the initial loop now also contains trefoil-like segments, as highlighted by the blue peaks in the cross-sectional maps, further indicating that  the character of the initial loop can sculpt the local geometric nature of the segments enclosed within the initial loop.}

\section{Conclusions}

This work developed a geometric and algebraic description of nematic disclination loops in three dimensions based on the Clifford algebra $\mathrm{Cl}(3,0)$~\cite{lounesto2001,hestenes2003}. This framework extends the standard topological classification of loops, which is encoded in the $\mathbb{Z}_4$ Janich index~\cite{janich1987,vcopar2013,vcopar2014}, by incorporating the full structure of the local defect profile and its transport along the loop. In particular, the Clifford formulation provides a compact representation of the SU(2) holonomy associated with the defect profile. This formulation allows the loop geometry to be systematically decomposed into distinct contributions, such as torsional self-twist and discrete profile transformations. This formulation can serve as a useful framework for analysing defect configurations in nematic liquid crystals in 3D.

We have also presented numerical simulations of nematic dynamics, showing that the insertion of a defect loop generically leads to the formation of transient disclination structures. A ``topological blob'' initially grows rapidly, reflecting the nucleation and proliferation of disclination segments, and subsequently decays on longer timescales towards a defect-free ordered state, with an apparent exponent of $t^{-1/2}$. 
This three-stages evolution appears robust for all studied configurations.

The main result reported here is that the early-time dynamics of transient structures retain clear signatures of the loop geometry, as encoded in the Clifford algebra description. In particular, loops where the defect profile is self-twisted lead to the nucleation of linked disclination loops. The number of links is determined by a simple arithmetic relation with the imposed twist ratio. In contrast, loops characterised by discrete profile changes, such as PMPM configurations, predominantly generate threaded defect structures without explicit topological linking, {providing an analogue of threadings associated with polymer melts~\cite{michieletto2014,smrek2016,smerk2020}}. These findings establish a direct connection between the geometric SU(2) holonomy of a disclination loop and the topology of the defect structures that emerge during its evolution.

A natural question is whether the transient defect structures identified here can be stabilised, for instance through confinement, patterned boundary conditions, activity~\cite{negro2025,amey2026,Ruske2021}, or the introduction of impurities, thereby enabling the design of tunable topological states. More broadly, it would be interesting to extend this approach to systems with more complex order parameter manifolds, such as cholesteric~\cite{machon2017contact,han2022uniaxial,beller2014,johnson2025,henrich2010,henrich2011}, and smectic liquid crystals~\cite{machon2019aspects,severino2025dislocations}, and to non-circular and knotted initial loops, where writhe and self-linking interact with the $\mathrm{SU}(2)$ profile holonomy during relaxation.

{\it Data availability statement:} The data that support the findings of this study are available upon reasonable request from the authors.

\bibliographystyle{eplbib.bst} 
\bibliography{bib.bib}
\clearpage
\includepdf[pages=-]{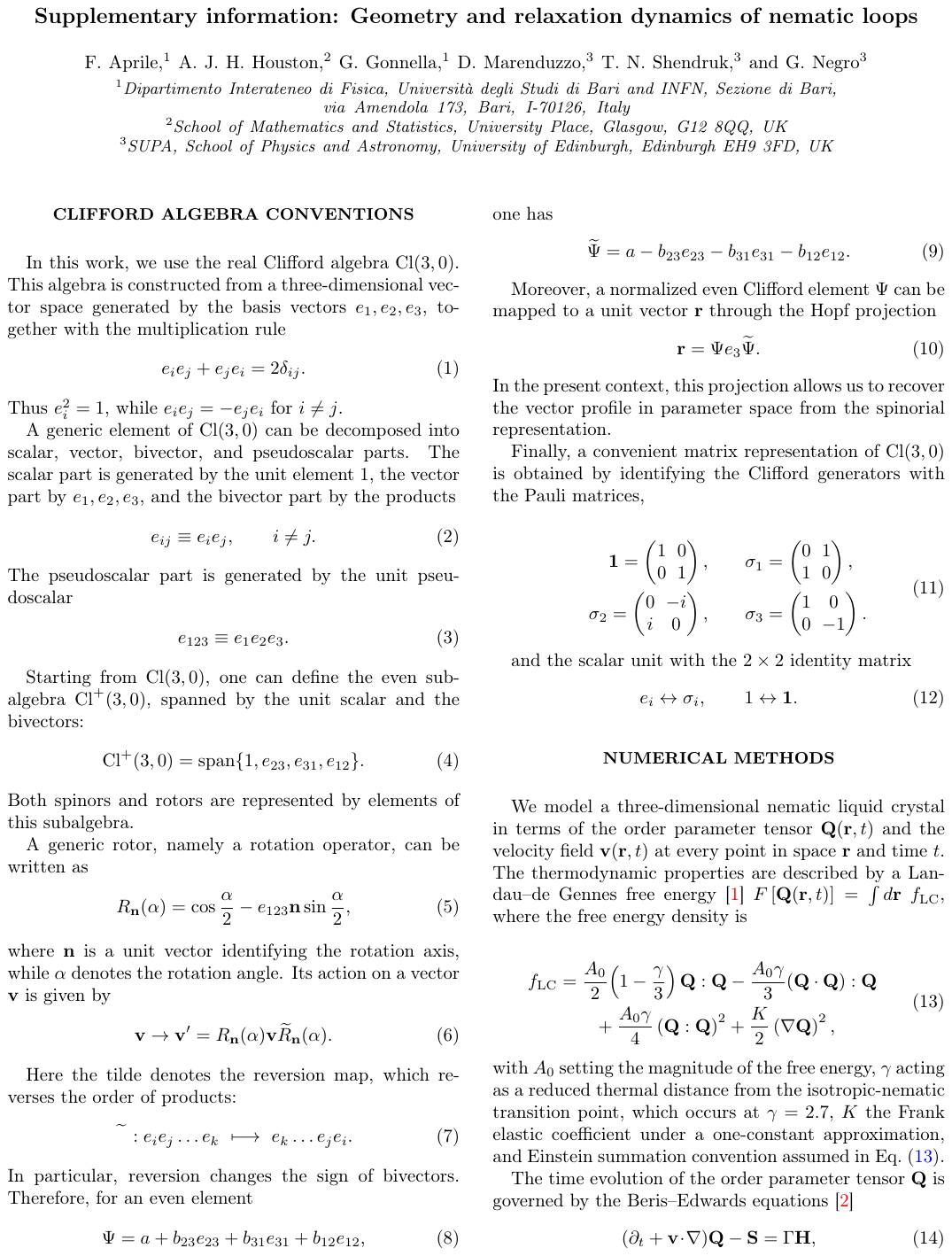}
\end{document}


\author{F. Aprile}
\affiliation{Dipartimento Interateneo di Fisica, Università degli Studi di Bari and INFN, Sezione di Bari,\\ via Amendola 173, Bari, I-70126, Italy}
\author{  A. J. H. Houston}
\affiliation{ School of Mathematics and Statistics, University Place, Glasgow, G12 8QQ, UK\\
}
\author{ G. Gonnella}
\affiliation{Dipartimento Interateneo di Fisica, Università degli Studi di Bari and INFN, Sezione di Bari,\\ via Amendola 173, Bari, I-70126, Italy}
\author{ D. Marenduzzo}
\affiliation{ SUPA, School of Physics and Astronomy, University of Edinburgh, Edinburgh EH9 3FD, UK}
\author{ T.~N. Shendruk}
\affiliation{ SUPA, School of Physics and Astronomy, University of Edinburgh, Edinburgh EH9 3FD, UK}
\author{ G. Negro}
\affiliation{ SUPA, School of Physics and Astronomy, University of Edinburgh, Edinburgh EH9 3FD, UK}

\title{Supplementary information: Geometry and relaxation dynamics of nematic loops}
\maketitle

\section{Clifford algebra conventions}

In this work, we use the real Clifford algebra \(\mathrm{Cl}(3,0)\). This algebra is constructed from a three-dimensional vector space generated by the basis vectors \(e_1,e_2,e_3\), together with the multiplication rule
\begin{equation}
    e_i e_j + e_j e_i = 2\delta_{ij}.
\end{equation}
Thus \(e_i^2=1\), while \(e_i e_j=-e_j e_i\) for \(i\neq j\).

A generic element of \(\mathrm{Cl}(3,0)\) can be decomposed into scalar, vector, bivector, and pseudoscalar parts. The scalar part is generated by the unit element \(1\), the vector part by \(e_1,e_2,e_3\), and the bivector part by the products
\begin{equation}
    e_{ij}\equiv e_i e_j, \qquad i\neq j .
\end{equation}
The pseudoscalar part is generated by the unit pseudoscalar
\begin{equation}
    e_{123}\equiv e_1 e_2 e_3 .
\end{equation}

Starting from \(\mathrm{Cl}(3,0)\), one can define the even subalgebra \(\mathrm{Cl}^+(3,0)\), spanned by the unit scalar and the bivectors:
\begin{equation}
    \mathrm{Cl}^+(3,0)=\mathrm{span}\{1,e_{23},e_{31},e_{12}\}.
\end{equation}
Both spinors and rotors are represented by elements of this subalgebra.

A generic rotor, namely a rotation operator, can be written as
\begin{equation}
    R_{\mathbf{n}}(\alpha)
    =
    \cos\frac{\alpha}{2}
    -
    e_{123}\mathbf{n}\sin\frac{\alpha}{2},
\end{equation}
where \(\mathbf{n}\) is a unit vector identifying the rotation axis, while \(\alpha\) denotes the rotation angle. Its action on a vector \(\mathbf{v}\) is given by
\begin{equation}
    \mathbf{v}\to \mathbf{v}'
    =
    R_{\mathbf{n}}(\alpha)\mathbf{v}\widetilde{R}_{\mathbf{n}}(\alpha).
\end{equation}

Here the tilde denotes the reversion map, which reverses the order of products:
\begin{equation}
    \widetilde{\phantom{A}}:
    e_i e_j \dots e_k
    \ \longmapsto\
    e_k \dots e_j e_i .
\end{equation}
In particular, reversion changes the sign of bivectors. Therefore, for an even element
\begin{equation}
    \Psi=a+b_{23}e_{23}+b_{31}e_{31}+b_{12}e_{12},
\end{equation}
one has
\begin{equation}
    \widetilde{\Psi}
    =
    a-b_{23}e_{23}-b_{31}e_{31}-b_{12}e_{12}.
\end{equation}

Moreover, a normalized even Clifford element \(\Psi\) can be mapped to a unit vector \(\mathbf{r}\) through the Hopf projection
\begin{equation}
    \mathbf{r}=\Psi e_3 \widetilde{\Psi}.
\end{equation}
In the present context, this projection allows us to recover the vector profile in parameter space from the spinorial representation.

Finally, a convenient matrix representation of \(\mathrm{Cl}(3,0)\) is obtained by identifying the Clifford generators with the Pauli matrices, 

\begin{equation}
\begin{aligned}
    \mathbf{1} =
    \begin{pmatrix}
    1 & 0\\
    0 & 1
    \end{pmatrix},
    \quad
    &\quad
    \sigma_1 =
    \begin{pmatrix}
    0 & 1\\
    1 & 0
    \end{pmatrix}, \\
    \quad
    \sigma_2 =
    \begin{pmatrix}
    0 & -i\\
    i & 0
    \end{pmatrix},
    \quad
    &\quad
    \sigma_3 =
    \begin{pmatrix}
    1 & 0\\
    0 & -1
    \end{pmatrix}.
\end{aligned}
\end{equation}

and the scalar unit with the \(2\times2\) identity matrix
\begin{equation}
    e_i\leftrightarrow \sigma_i,
    \qquad
    1\leftrightarrow \mathbf{1}.
\end{equation}

\section{Numerical methods}

We model a three-dimensional nematic liquid crystal in terms of the order parameter tensor $\mathbf{Q}(\mathbf{r},t)$ and the  velocity field $\mathbf{v}(\mathbf{r},t)$ at every point in space $\mathbf{r}$ and time $t$. 
The thermodynamic properties are described by a Landau–de Gennes free energy~\cite{degennes1993} \(F\left[\mathbf{Q}(\mathbf{r},t)\right] = \int d\mathbf{r} \ f_\mathrm{LC} \), where the free energy density is

\begin{equation}\label{FreeEnergy}
\begin{split}
    f_{\mathrm{LC}} &=
        \frac{A_0}{2}\!\left(1-\frac{\gamma}{3}\right) \mathbf{Q}:\mathbf{Q}
        -\frac{A_0\gamma}{3} (\mathbf{Q}\cdot\mathbf{Q}):\mathbf{Q} \\
        &\qquad +\frac{A_0\gamma}{4}\left(\mathbf{Q}:\mathbf{Q}\right)^2
        +\frac{K}{2}\left(\nabla \mathbf{Q}\right)^2,
\end{split}
\end{equation}
with \(A_0\) setting the magnitude of the free energy, \(\gamma\) acting as a reduced thermal distance from the isotropic-nematic transition point, which occurs at \(\gamma=2.7\), \(K\) the Frank elastic coefficient under a one-constant approximation, and Einstein summation convention assumed in Eq.~\eqref{FreeEnergy}. 

The time evolution of the order parameter tensor $\mathbf{Q}$ is governed by the Beris–Edwards equations~\cite{beris}
\begin{equation}\label{BEeq}
    (\partial_t + \mathbf{v}\!\cdot\!\nabla)\mathbf{Q} - \mathbf S = \Gamma \mathbf{H},
\end{equation}
where \(\Gamma\) is the rotational diffusion coefficient, $\mathbf{H}=-\frac{\delta f}{\delta \mathrm{Q}}$ denotes the molecular field, and \(\mathbf S\) describes the rotation of the nematic molecules induced by velocity gradients that depends on the tumbling parameter \(\xi\)~\cite{carenza2019}.

The fluid velocity evolves according to the incompressible Navier–Stokes equations,
\begin{align} \label{NSeq}
    \rho(\partial_t\mathbf{v}+\mathbf{v}\!\cdot\!\nabla\mathbf{v})
    &= -\nabla p + \nabla\!\cdot\!\left(\boldsymbol{\sigma}^{\mathrm{visc}}
    +\boldsymbol{\sigma}^{\mathrm{LC}}\right),\\
    \nabla\!\cdot\!\mathbf{v} &= 0,
\end{align}
where \(\rho\) is the mass density, \(p\) is the isotropic pressure and the total stress tensor $\boldsymbol{\sigma}$ is decomposed into viscous $\boldsymbol{\sigma}^\mathrm{visc}=\eta\nabla\cdot\left[\nabla\mathbf{v}+\nabla\mathbf{v}^\top\right]$  and liquid crystal $\boldsymbol{\sigma}^\mathrm{LC}$ contributions.

The systems analyzed in the main consist of a three-dimensional box of size \(L_x=L_y=L_z=64,,74,128\).

The director field for self-twisted loop configuration outside the toroidal region is initialized as 
\begin{align}
    n_x &= -\sin\!\left(-{\phi}/{2} + \tau(\theta)\right)\cos\theta, \\
    n_y &= -\sin\!\left(-{\phi}/{2} + \tau(\theta)\right)\sin\theta, \\
    n_z &= \cos\!\left(-{\phi}/{2} + \tau(\theta)\right),
\end{align}
where \(\phi\) is the angle parameterizing the minor radius of the torus and \(\theta\) parametrize the major radius of the torus. 

The director field for PM and PMPM loops inside the toroidal region is initialized as
\begin{align}
    n_x &= -\sin\!\left(\phi/2\right)\cos\!\left(\theta + \beta(\theta)\right), \\
    n_y &= -\sin\!\left(\phi/2\right)\sin\!\left(\theta + \beta(\theta)\right), \\
    n_z &= \cos\!\left(\phi/2\right).
\end{align}
Here \(\beta(\theta)\) controls the variation of the local profile along the loop. 
In particular, \(\beta(\theta)=\theta\) corresponds to a PM loop, while \(\beta(\theta)=-2\theta\) characterizes the PMPM defect loop.

In both cases outside the toroidal region the system is initialized as isotropic, namely the director field is normalized and orientated in a random direction.

Parameters used in all the studied configurations are \(\xi=1\), \(A_0=0.5\), \(K=0.01\), \(\Gamma=0.3375\), \(\gamma=3\), \(\eta = 5/3\), \(\rho = 2\). 
The time step \(\Delta t = 1\) and the spatial discretisation \(\Delta x = 1\) set the time and space units.

To identify the defect structures, we employ two strategies. First, the maximum eigenvalue of the nematic tensor \(\mathbf{Q}(\mathbf{r},t)\) corresponds to the local scalar order parameter \(q(\mathbf{r},t)\) and we visualize the isosurfaces of $q = 0.3$. Second, we utilize the disclination tensor~\cite{schimming2022,head2024,head2024b,negro2025}
\begin{equation}
    D_{ij} = \varepsilon_{i\mu\nu}\,\varepsilon_{jlk}\,\partial_l Q_{\mu \alpha}\,\partial_k Q_{\nu \alpha},
\end{equation}
constructed from the spatial derivatives of the nematic order parameter tensor. The disclination tensor admits the decomposition
\begin{equation}
    \mathbf{D} = s(\mathbf{r},t)\, \mathbf{\Omega} \otimes \mathbf{T},
\end{equation}
where \( s(\mathbf{r},t) \) is a positive scalar field that is maximum at the disclination cores, \( \mathbf{\Omega} \) is the local winding vector, and \( \mathbf{T} \) is the unit tangent vector to the defect line.
Defects are identified as isosurfaces of the scalar field \( s \), using a threshold value \( s_{\mathrm{cut}} = 0.075 \). 

The trace of the disclination tensor \(\mathrm{Tr}\,\mathbf{D} = s\cos\beta \) represents a defect-density field, where, as before, \( \beta \) denotes the angle between the winding vector \( \mathbf{\Omega} \) and the tangent vector to the loop \( \mathbf{T} \). 
This allows us to distinguish between regions
with different local defect character. In particular, positive values of \(\cos\beta\) correspond to comet-like segments, negative values correspond
to trefoil-like segments, and \(\cos\beta=0\) identifies twist-like segments. In order to capture the local defects geometry induced by the presence of the initial loop in the system,  the  defect-density field was averaged across a thin slab of thickness equal to the diameter of the little circle of the torus, across the direction perpendicular to the initial loop.

\section{Supplementary figures}
\begin{figure}[t!]
    \centering
    \includegraphics[width=\columnwidth]{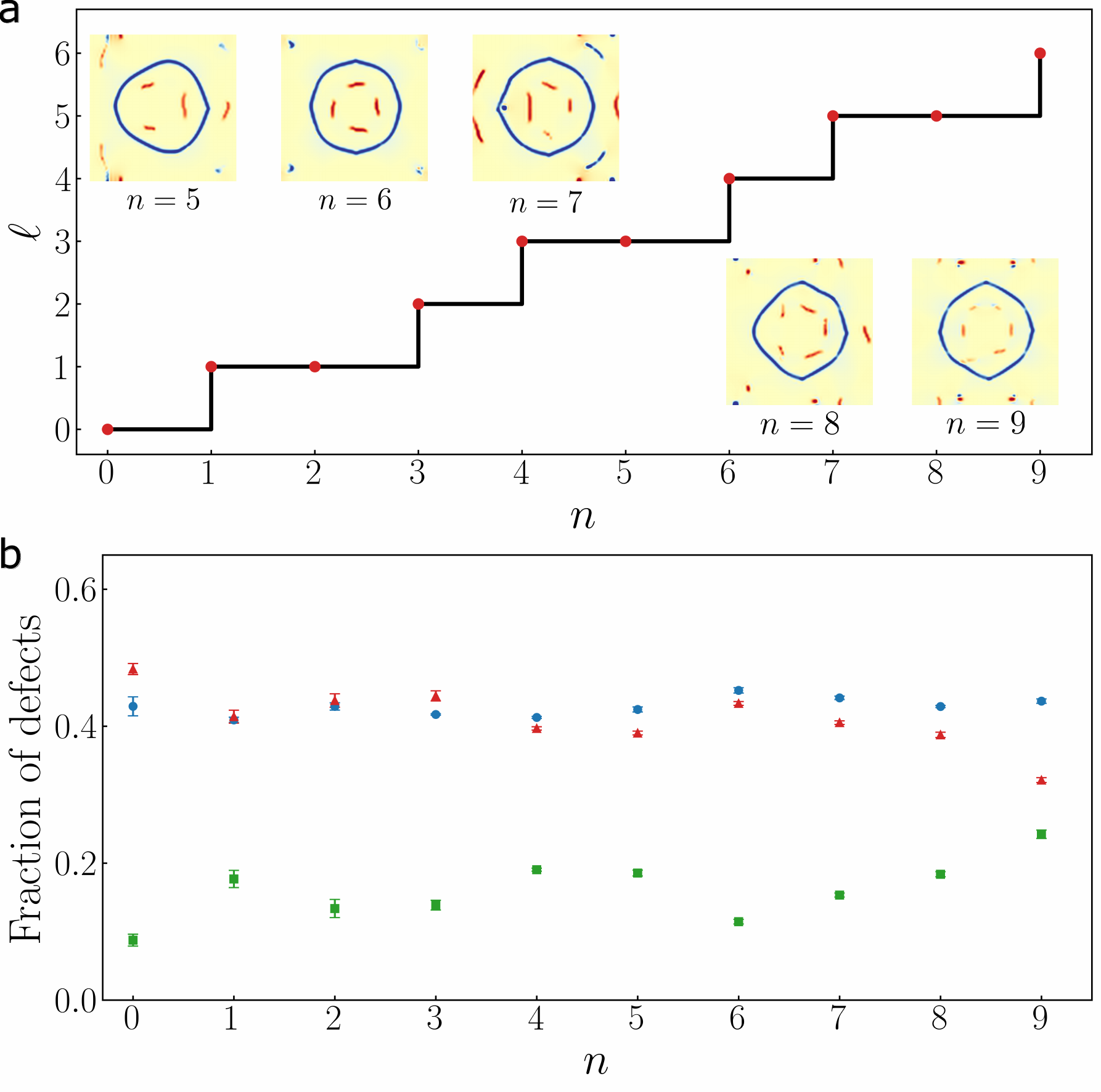}
    \caption{
    \textbf{Characterization of self-twisted wedge loops.}
    (a) Number of formed links for initial wedge-type \( -1/2 \) profiles with
    self-twist \( \tau = n\theta/3 \), where \( n \in \mathbb{Z} \) is the twist-rate index.
    The black line denotes the theoretical prediction, while red dots denote the
    values of \( n \) for which the theoretical relation was checked.
    Insets show cross-sectional maps of \( s\cos\beta \) for
    \( n = 5, 6, 7, 8, 9 \), averaged along the direction perpendicular to the initial loop.
    (b) Fraction of defect for all points within the system for the same initial configurations.
    A point is classified as trefoil for \( \cos\beta \in [-1,-0.5) \) (blue circles),
    twist for \( \cos\beta \in [-0.5,0.5] \) (green squares), and comet for
    \( \cos\beta \in (0.5,1] \) (red triangles).
    Reported values are averages over six independent simulations;
    error bars represent the standard error of the mean.}
    \label{fig4}
\end{figure}

Here we present additional results on the characterization of self-twisted wedge loops. Fig.~\ref{fig4}a shows the relation between the number of links formed during the relaxation of defect profiles with self-twist \(\tau=n\theta/3\) and the twisting ratio \(n\). The insets show additional cross-sectional maps of \(s\cos{\beta}\). In Fig.~\ref{fig4}b, we characterize the local geometry of the defect structures as a function of the twisting ratio of the initial loop. The analyzed defect structures consistently exhibit a prevalence of trefoil and comet profiles, whereas local twist profiles are less frequent.

\bibliography{bib}